\documentclass[twocolumn,showpacs,superscriptaddress,amsmath,amssymb]{revtex4}
\usepackage{graphicx}
\usepackage{dcolumn}
\usepackage{bm}
\usepackage{color}
\usepackage{float}

\def\bd{
\begin{document}} \def\ed{\end{document}}
\def\bmp{\begin{minipage}} \def\emp{\end{minipage}}
\def\bcc{\begin{center}} \def\ecc{\end{center}}     \def\npg{\newpage}
\def\beq{\begin{equation}} \def\eeq{\end{equation}} \def\hph{\hphantom}
\def\be{\begin{equation}} \def\ee{\end{equation}} \def\r#1{$^{[#1]}$}
\def\n{\noindent} \def\ni{\noindent} \def\pa{\parindent}
\def\hs{\hskip} \def\vs{\vskip} \def\hf{\hfill} \def\ej{\vfill\eject}
\def\cl{\centerline} \def\ob{\obeylines}  \def\ls{\leftskip}
\def\underbar#1{$\setbox0=\hbox{#1} \dp0=1.5pt \mathsurround=0pt
   \underline{\box0}$}   \def\ub{\underbar}    \def\ul{\underline}
\def\f{\left} \def\g{\right} \def\e{{\rm e}} \def\o{\over} \def\d{{\rm d}}
\def\vf{\varphi} \def\pl{\partial} \def\cov{{\rm cov}} \def\ch{{\rm ch}}
\def\la{\langle} \def\ra{\rangle} \def\EE{e$^+$e$^-$} \def\pt{p_{\rm t}}
\def\pti{p_{{\rm t},i}} \def\vti{v_{{\rm t},i}}
\def\ptj{p_{{\rm t},j}}\def\Pt{P_{\rm t}} \def\vt{v_{\rm t}}
\def\YT{Y_{{\rm T}}} \def\yT{y_{{\rm T}}}
\def\yTi{y_{{\rm T},i}}

\def\bitz{\begin{itemize}} \def\eitz{\end{itemize}}
\def\btbl{\begin{tabular}} \def\etbl{\end{tabular}}
\def\btbb{\begin{tabbing}} \def\etbb{\end{tabbing}}
\def\beqar{\begin{eqnarray}} \def\eeqar{\end{eqnarray}}
\def\\{\hfill\break} \def\dit{\item{-}} \def\i{\item}
\def\bbb{\begin{thebibliography}{9} \itemsep=-1mm}
\def\ebb{\end{thebibliography}} \def\bb{\bibitem}
\def\bpic{\begin{picture}(260,240)} \def\epic{\end{picture}}
\def\akgt{\cl{\bf ACKNOWLEDGMENTS}}
\def\fgn{\noindent{\bf\large\bf figure captions}}
\def\m1{\langle N_p\rangle} \def\u2{\langle N_{\bar p}\rangle} \def\Nap{N_{\bar
p}}
\def\lan{\langle}
\def\ran{\rangle}
\def\p{\pi}
\def\ifmath#1{\relax\ifmmode #1\else $#1$\fi}%
\def\rc{\ifmath{{\mathrm{c}}}}
\def\cut{\ifmath{{\mathrm{cut}}}}
\def\rF{\ifmath{{\mathrm{F}}}}
\def\rK{\ifmath{{\mathrm{K}}}}
\def\rp{\ifmath{{\mathrm{p}}}}
\def\rt{\ifmath{{\mathrm{t}}}}
\def\LAB{\ifmath{{\mathrm{LAB}}}}
\def\cut{\ifmath{{\mathrm{cut}}}}
\def\beq{\begin{equation}}
\def\eeq{\end{equation}}

\newcommand{\cinst}[2]{$^{\mathrm{#1}}$~#2\par}
\newcommand{\crefi}[1]{$^{\mathrm{#1}}$}
\newcommand{\crefii}[2]{$^{\mathrm{#1,#2}}$}
\newcommand{\crefiii}[3]{$^{\mathrm{#1,#2,#3}}$}
\newcommand{\HRule}{\rule{0.5\linewidth}{0.5mm}}

\bd
\title{ Measurement methods of radial flow in relativistic heavy-ion collisions}

\author{Peng Yang}
\affiliation{Key Laboratory of Quark and Lepton Physics (MOE), Institute of Particle Physics, Central China Normal University, Wuhan 430079, China}
\author{Lin Li}
\affiliation{School of Science, Wuhan University of Technology, Wuhan 430070, China}
\author{Yu Zhou}
\affiliation{Department of Mathematics, University of California, Los Angeles, California 90095, USA}
\author{Zhiming Li}
\affiliation{Key Laboratory of Quark and Lepton Physics (MOE), Institute of Particle Physics, Central China Normal University, Wuhan 430079, China}
\author{Mingmei Xu}
\affiliation{Key Laboratory of Quark and Lepton Physics (MOE), Institute of Particle Physics, Central China Normal University, Wuhan 430079, China}
\author{Yeyin Zhao}
\affiliation{Key Laboratory of Quark and Lepton Physics (MOE), Institute of Particle Physics, Central China Normal University, Wuhan 430079, China}
\author{Yuanfang Wu}\email{wuyf@mail.ccnu.edu.cn}
\affiliation{Key Laboratory of Quark and Lepton Physics (MOE), Institute of Particle Physics, Central China Normal University, Wuhan 430079, China}

\begin{abstract}

Radial flow can be directly extracted from the azimuthal distribution of mean transverse rapidity. We apply the event-plane method and the two-particle correlation method to estimate the anisotropic Fourier coefficient of the azimuthal distribution of mean transverse rapidity. Using the event sample generated by a multiphase transport model with string melting, we show that both methods are effective. For the two-particle correlation method to be reliable, the mean number of particles in an azimuthal bin must be above a certain threshold. Using these two methods, anisotropic radial flow can be estimated in a model-independent way in relativistic heavy-ion collisions.

\end{abstract}


\maketitle
\section{Introduction}

Collectivity, or flow, is one of the main characteristics of the newly formed quark-gluon plasma in relativistic heavy-ion collisions~\cite{lab1,lab2,lab3,lab4}. Traditionally, we have observed radial flow, direct flow, elliptic flow, triangular flow, and so on. Radial flow and elliptic flow are the two prominent classes among them.

Elliptic flow is defined as the second Fourier component of the azimuthal multiplicity distribution. It is generated by the initial geometric asymmetry of noncentral collisions where the overlap of two incident nuclei is of an almond shape in the transverse coordinate plane. The minor axis of the overlap is on the reaction plane, which is spanned by the vector of impact parameter and beam direction~\cite{lab1,lab4,lab5}. This initial geometric asymmetry leads to a larger density gradient along the in-plane direction, i.e., the anisotropic distribution of final-state particles in momentum space.
Therefore, elliptic flow provides information about initial conditions and system properties~\cite{lab2,lab6}.

Radial flow is originally deduced from an analysis of transverse momentum spectra in central collisions~\cite{lab1,lab4,lab7}. Later, it is generalized into two parameters: isotropic radial velocity and anisotropic radial velocity~\cite{lab8,lab9,lab10,lab11,lab12}. The isotropic radial velocity features the isotropic transverse expansion of the source at kinetic freeze-out. The anisotropic radial velocity presents the difference of the radial flow between the in-plane direction and the out-of-plane direction and arises in noncentral collisions. 


The interplay of radial expansion and elliptic flow results in what we observe as the particle mass splitting of the differential elliptic flow~\cite{lab3,lab4,lab6}. Namely, the   heavier particles show smaller elliptic flow values. This mass ordering of elliptic flow has been well understood by hydrodynamics with a set of kinetic freeze-out constraints, i.e., radial flow, temperature, and source deformation~\cite{lab13,lab14}.

In addition, momentum transfer due to viscosity is proportional to the first derivative of the velocity in hydrodynamics~\cite{lab15,lab16}. Bulk viscosity is associated with isotropic transverse velocity, whereas shear viscosity is associated with the anisotropic transverse velocity. The proportion constants are equal to the bulk and shear viscosities, respectively. Therefore, the determination of anisotropic radial flow is essential for hydrodynamic calculations and for measuring the shear viscosity in relativistic heavy-ion collisions~\cite{lab17,lab18,lab19}.

Conventionally, anisotropic radial flow is extracted by fitting the transverse momentum spectrum of particles with the blast-wave parametrization~\cite{lab20,lab21}. To minimize $\chi^2$ of the fitting, the spectra of all particle species are fitted simultaneously. For the relativistic heavy-ion collisions at the Brookhaven National Laboratory (BNL) Alternating Gradient Synchrotron, the BNL Relativistic Heavy Ion Collider (RHIC), and the CERN Large Hadron Collider energies, the fitting results have all been presented~\cite{lab22,lab23,lab24,lab25}. However, these results are obviously model dependent. Therefore, a model-independent measurement is called for.

In order to measure radial flow, we had introduced the mean-transverse rapidity (MTR) and its azimuthal distribution, replacing velocity with rapidity in the original definition of radial flow~\cite{lab12,lab15,lab26}. The MTR is averaged over the number of particles in an azimuthal angle bin. It presents the raw kinetic expansion in a specified azimuthal direction.

The azimuthal distribution of the MTR presents the transverse expansion of the source at kinetic freeze-out~\cite{lab15,lab27,lab28}. The second Fourier coefficient of this distribution features the anisotropic kinetic expansion and is consistent with the anisotropic radial flow extracted from the blast-wave parametrization~\cite{lab12}. 


The aim of this paper is to estimate the anisotropic radial flow in relativistic heavy-ion collisions. This estimation is similar to how we estimate elliptic flow where the reaction plane cannot be measured directly due to random fluctuations of the impact parameter vector and is instead estimated from the reconstructed particles. This estimation method is the so-called event-plane method (EPM)~\cite{lab29,lab30,lab31}. Later, it is found that the two-particle correlation method (TPCM) can also be used in place of the EPM to estimate the elliptic flow~\cite{lab32,lab33,lab34,lab35,lab36}. 

In this paper, we first describe the EPM and the TPCM that are designed to estimate the anisotropic coefficients of azimuthal distributions for both total-transverse rapidity (TTR) and MTR in Sec. II. Then, we demonstrate the effectivity of both methods using the event sample generated by the AMPT model with string melting in Sec. III. The limitation of the TPCM is discussed in Sec. IV. Finally, summary and conclusions are presented in Sec. V.

 
\section{Measurement methods}

Transverse rapidity is defined as~\cite{lab2,lab12,lab15,lab26}

\begin{equation}
\label{eq3}  y_{T}=\ln\left(\frac{m_{T}+p_{T}}{m_{0}}\right),
\end{equation}
\noindent where $m_{0}$ is the particle mass in the rest frame, $p_{T}$ is transverse momentum, and $m_{T}=\sqrt{m_{0}^2+p_{T}^2}$ is the transverse mass. 

The TTR of the \emph{m}th azimuthal bin is the summation of all  particles' transverse rapidities in an event, and its average is~\cite{lab12,lab15,lab26} as follows:

\begin{equation}\label{YT}
\la Y_{T}(\phi_m-\Psi_r)\ra=\frac{1}{N_{\rm event}}\sum_{k=1}^{N_{\rm event}}\sum_{i=1}^{N_m^k}y_{T,i}^k(\phi_{m}-\Psi_{r}).
\end{equation}

\noindent It measures total transverse expansion in a specified azimuthal bin $(\phi_m-\Psi_r)$, where $\Psi_{r}$ is the reaction plane angle. $\phi_{m}$ is the \emph{m}th azimuthal angle bin. The bin width is $2\pi/N_{\rm bin}$. $N_{\rm bin}$ is the total number of bins. $y_{T,i}^k(\phi_m-\Psi_r)$ is the transverse rapidity of the \emph{i}th particle in the \emph{k}th event and \emph{m}th azimuthal angle bin. $N_m^k$ is the total number of particles in the \emph{k}th event and the \emph{m}th azimuthal angle bin. $N_{\rm event}$ is the total number of events.  

The MTR of the \emph{m}th azimuthal bin is defined as

\begin{equation}\label{yT}
\la\la\yT(\phi_m-\Psi_r)\ra\ra=\frac{1}{N_{\rm event}}\sum_{k=1}^{N_{\rm event}}\frac{1}{N_m^k}\sum_{i=1}^{N_m^k}y_{T,i}^k(\phi_m-\Psi_{r}),
\end{equation}

\noindent where $\la\la\cdots\ra\ra$ is first averaged over the number of particles in the \emph{k}th event and the \emph{m}th azimuthal angle bin $N_m^k$, and then over the total number of events. It measures the raw kinetic expansion in a specified azimuthal direction. The contribution of the number of particles is removed by the first average.

The Fourier expansions of azimuthal TTR and MTR distributions are as follows:

\begin{equation}\label{Fourier-YT}
 \begin{split}
&\frac{d\la Y_{T}(\phi-\Psi_r)\ra}{ d(\phi-\Psi_r)}\\ &\ =v_0(Y_T)\left( 1+\sum_{n=1}^{\infty}2v_{n}(\YT)\cos\left[n\left(\phi-\Psi_{r}\right)\right]\right),
 \end{split}
\end{equation}

\noindent and

\begin{equation}\label{Fourier-yT}
\begin{split}
 &\frac{d\la\la y_{T}(\phi-\Psi_r)\ra\ra}{d(\phi-\Psi_r)}\\ &\ =v_0(y_T)\left( 1+\sum_{n=1}^{\infty}2v_{n}( y_{T})\cos\left[n\left(\phi-\Psi_{r}\right)\right]\right),
 \end{split}
\end{equation}

\noindent where azimuthal-angle-independent $v_{0}(Y_{T})$ and $v_{0}(y_{T})$ are  isotropic TTR flow and radial flow, respectively. The second Fourier coefficients $v_{2}(Y_{T})$ and $v_{2}(y_{T})$ are defined as anisotropic TTR flow and radial flow, respectively~\cite{lab12,lab15,lab26}. 


They are similar to the mean multiplicity in the \emph{m}th azimuthal angle bin, i.e.,

\begin{equation}\label{N}
\la N(\phi_m-\Psi_r)\ra=\frac{1}{N_{\rm event}}\sum_{k=1}^{N_{\rm event}}N^k(\phi_{m}-\Psi_{r}),
\end{equation}

\noindent where $N^k(\phi_m-\Psi_r)$ is the number of particles in the \emph{k}th event and the $(\phi_m-\Psi_r)$ direction. Its azimuthal distribution in Fourier expansion is as follows:

\begin{equation}\label{Fourier-N}
\begin{split}
&\frac{d\la N(\phi-\Psi_r)\ra}{d(\phi-\Psi_r)}\\ &\ =v_0(N)\left( 1 +\sum_{n=1}^{\infty}2v_{n}(N)\cos\left[n\left(\phi-\Psi_{r}\right)\right]\right),
 \end{split}
\end{equation}

\noindent where $v_{0}(N)$ is azimuthal-angle independent. $v_{2}(N)$ is the elliptic flow, which is estimated, in practice, by the EPM and the TPCM. In the following, we apply these two methods to estimate $v_{2}(Y_{T})$ and $v_{2}(y_{T})$.

\subsection{The event-plane method}

In nuclear-nuclear collisions, the azimuthal angle of the final-state particle is measured with respect to the reaction plane. However, the angle of reaction plane is unknown in experiments and fluctuates from event to event. Usually, a reconstructed event plane is considered as a substitute~\cite{lab29}, i.e., $\Psi_{n}$ of the \emph{n}th coefficient is given by

\begin{equation}\label{eq9} 
\Psi_{n}=\left.\left(\tan^{-1}\frac{Q_{n,y}}{Q_{n,x}}\right)\middle/n\right.,
\end{equation}

\noindent where $Q_{n}$ is the flow vector and defined as, 

\begin{equation}
\label{eq7} Q_{n,x}=\sum_{i=1}^{M}\omega_{i}\cos(n\phi_{i})=Q_{n}\cos(n\Psi_{n}),
\end{equation}

\begin{equation}
\label{eq8} Q_{n,y}=\sum_{i=1}^{M}\omega_{i}\sin(n\phi_{i})=Q_{n}\sin(n\Psi_{n}).
\end{equation}

\noindent \emph{M} is the number of particles used in the event-plane determination. $\phi_{i}$ and $\omega_{i}$ are the azimuthal angle and the weight of the \emph{i}th particle. 

A general Fourier expansion with respect to the event-plane ($\Psi_n$) can be expressed as
\begin{equation}\label{EP-Fourier} 
\begin{split}
&\frac{d\la \omega N(\phi-\Psi_n)\ra}{d(\phi-\Psi_n)}\\ &\ =v_{0}(\omega N)\left(1+\sum_{n=1}^{\infty}2v_n^{\rm obs}(\omega N)\cos\left[n\left(\phi-\Psi_{n}\right)\right]\right),
\end{split}
\end{equation}
\noindent where weight $\omega$ changes with the observable~\cite{lab29}, such as different kinds of flow defined in Eqs.~(\ref{Fourier-YT}), (\ref{Fourier-yT}), and (\ref{Fourier-N}). 

For multiplicity, the weight is unity, and Eq.~(\ref{EP-Fourier}) corresponds to
Eq.~(\ref{Fourier-N}). Its $v_{2}^{\rm obs}(N)$ is as follows:

\begin{equation}\label{v2RP-N} 
v_{2}^{\rm obs}\left( N \right) = \frac{1}{N_{\rm event}} \sum_{k=1}^{N_{\rm event}} \left( \frac{1}{N^k} \sum_{i=1}^{N^k}\cos\left[ 2\left( \phi_{i}^k-\Psi _{2} \right) \right]\right),
\end{equation}

\noindent where $\Psi_{2}$ is the event-plane angle of the second harmonic and $N^k$ is the total number of particles in the \emph{k}th event.

For the TTR and MTR, their weights are $\omega_i=y_{T,i}$ and 
$\omega_i=\left. y_{T,i,m}\middle/N_m \right.$, respectively. Their $v_{2}^{\rm obs}(Y_T)$ and $v_{2}^{\rm obs}(y_T)$ are as follows:

\begin{equation}\label{v2RP-YT} 
\begin{split}
v_{2}^{\rm obs}\left ( Y_{T} \right )=&\frac{1}{N_{\rm event}}\sum_{k=1}^{N_{\rm event}}\\   &\left ( \frac{1}{\sum_{i=1}^{N^k}y_{T,i}^k} \sum_{i=1}^{N^k}y_{T,i}^k\cos\left [ 2\left ( \phi _{i}^k-\Psi _{2} \right ) \right ]\right ),
\end{split}
\end{equation}

\noindent and 

\begin{equation}\label{v2RP-yT}
\begin{split}
v_{2}^{\rm obs} \left( y_T \right)=&\frac{1}{N_{\rm event}}\sum_{k=1}^{N_{\rm event}} \left( \frac{1}{\sum_{m=1}^{N_{\rm bin}} \frac{1}{N_{m}^k}\sum_{k=1}^{N_m^k} y_{T,i,m}^k} \right.\\   
&\left.\sum_{m=1}^{N_{\rm bin}}\frac{1}{N_m^k}\sum_{k=1}^{N_m^k}y_{T,i,m}^k\cos\left[ 2\left( \phi _{i,m}^k-\Psi _{2} \right) \right] \right).
\end{split}
\end{equation}

With finite multiplicity,  $v_{2}^{\rm obs}(N)$ has to be corrected for the event-plane resolution~\cite{lab4,lab29},  i.e., 

\begin{equation}
\rm Re_{2}=\left \langle \cos\left [ 2(\Psi_{2}-\Psi_{r}) \right ] \right \rangle,
\end{equation}

\noindent which is obtained by the iteration of sub-event-plane resolution. The subevent is usually constructed in a longitudinal rapidity window that is different from the particles of interest~\cite{lab23,lab37}. Therefore, the elliptic flow is as follows:

\begin{equation}
\label{v2-N} v_{2} (N)=\left. v_{2}^{\rm obs}(N)\middle/\rm Re_{2} \right..
\end{equation}

\noindent After the same corrections for both $v_{2}^{\rm obs}(Y_T)$  and $v_{2}^{\rm obs}(y_T)$, $v_{2}(Y_T)$ and $v_{2}(y_T)$ are as follows:

\begin{equation}
\label{v2-YT} v_{2} (Y_T)=\left. v_{2}^{\rm obs}(Y_T)\middle/\rm Re_{2} \right.,
\end{equation}

\noindent and

\begin{equation}
\label{v2-yT} v_{2}(y_T)=\left. v_{2}^{\rm obs}(y_T)\middle/\rm Re_{2} \right..
\end{equation}

\subsection{The two-particle correlation method}

The coefficients of Fourier expansion Eq.~(\ref{Fourier-N})  can also be estimated by the TPCM~\cite{lab30}, i.e.,

\begin{equation}\label{v22}
\left \langle \cos\left [ n\left ( \phi _{1}-\phi _{2} \right ) \right ] \right \rangle=\left \langle e^{in\left ( \phi _{1}-\phi _{2} \right )} \right \rangle=\left \langle v_{n}^{2} \right \rangle+\delta _{n}.
\end{equation}

\noindent Here a cross term that depends both on $v_n$ and on $\delta_n$ is neglected as it is very small in most cases of interest. $\delta _{2}$ represents the so-called nonflow contribution and is irrelevant to initial geometry. All correlations are first averaged over all particle pairs in a given event and then over all events. The latter average involves weight depending on event. For convenience, single-event average two-particle azimuthal correlations is defined as

\begin{equation}
\label{v22single}\left \langle 2 \right \rangle^k\equiv \frac{1}{\sum_{i,j=1,i\neq j}^{N^k}\omega _i^k\omega _j^k}\sum_{i,j=1,i\neq j}^{N^k}\omega _i^k\omega _j^ke^{in\left ( \phi _i^k-\phi _j^k \right )}.
\end{equation}

\noindent Its average over all events is as follows:

\begin{equation}
\begin{split}
\label{v22average}\left \langle \left \langle 2 \right \rangle \right \rangle&\equiv\frac{\sum_{k=1}^{N_{\rm event}} \sum_{i,j=1,i\neq j}^{N^k} \omega _i^k\omega _j^k \left \langle 2 \right \rangle^k}{\sum_{k=1}^{N_{\rm event}} \sum_{i,j=1,i\neq j}^{N^k} \omega _i^k\omega _j^k}\\&=\frac{\sum_{k=1}^{N_{\rm event}} \sum_{i,j=1,i\neq j}^{N^k} \omega _i^k\omega _j^ke^{in\left ( \phi _i^k-\phi _j^k \right ) }}{\sum_{k=1}^{N_{\rm event}} \sum_{i,j=1,i\neq j}^{N^k} \omega _i^k\omega _j^k }.
\end{split}
\end{equation}

For multiplicity and TTR, their weights are $\omega_i=1$
and $\omega_i=y_{T,i}$, respectively. So their  $\left \langle \left \langle 2 \right \rangle \right \rangle$ are 

\begin{equation}\label{22-N}
\left \langle \left \langle 2 \right \rangle \right \rangle_N =\frac{\sum_{k=1}^{N_{\rm event}} \sum_{i,j,i\neq j}^{N^k} e^{in\left ( \phi _i^k-\phi _j^k \right ) }}{\sum_{k=1}^{N_{\rm event}}N^k\left ( N^k-1 \right )},
\end{equation}

\noindent and

\begin{equation}\label{22-YT}
\left \langle \left \langle 2 \right \rangle \right \rangle_{Y_T }=\frac{\sum_{k=1}^{N_{\rm event}}\sum_{i,j,i\neq j}^{N^k} y_{T,i}^ky_{T,j}^ke^{in\left ( \phi _i^k-\phi _j^k \right ) }}{\sum_{k=1}^{N_{\rm event}}\sum_{i,j,i\neq j}^{N^k}y_{T,i}^ky_{T,j}^k}.
\end{equation}

Elliptic flow can be estimated from two-particle correlations by~\cite{lab30,lab31}

\begin{equation}\label{v22-N}
v_{2} (N) =\sqrt{\left \langle \left \langle 2 \right \rangle \right \rangle_N}.
\end{equation}

Similarly, corresponding anisotropic coefficient of TTR can be estimated by,

\begin{equation}\label{v22-YT}
v_{2}(Y_T) =\sqrt{\left \langle \left \langle 2 \right \rangle \right \rangle_{Y_T}}.
\end{equation}

\noindent  Equation~(\ref{v22-YT}) is obtained in the same way as Eq.~(\ref{v22-N}) except for the nonunit weight factor. 

For MTR, its weight is $\omega_i=\left. y_{T,i,m}\middle/N_m \right.$. The two correlated particles could come from the same azimuthal angle bin or from different bins. If they come from different bins, the single-event average two-particle azimuthal correlations is as follows:

\begin{equation}\label{v2obs-Yt}
\begin{split}
\langle 2\rangle_{\rm d}^k=&\sum_{m,n=1,m\neq n}^{N_{\rm bin}}\left( \frac{1}{N_m^kN_n^k}
\right.\\ 
&\left.\sum_{i=1}^{N_m^k}\sum_{j=1}^{N_n^k}y_{T,i,m}^ky_{T,j,n}^ke^{in (\phi_{i,m}^k-\phi_{j,n}^k) }\right).
\end{split}
\end{equation}

\noindent Here $N_m^k$ and $N_n^k$ are the numbers of particles in the \emph{m}th and \emph{n}th bin, respectively. If they come from the same bin, the single-event average  two-particle azimuthal correlations are as follows:

\begin{equation}\label{v2obs-yt}
\begin{split}
\langle 2\rangle_{\rm s }^k=&\sum_{m=1}^{N_{\rm bin}} \left( \frac{1}{N_m^k (N_m^k-1)} \right.\\ 
&\left.\sum_{i,j=1,i\neq j}^{N_m^k} y_{T,i,m}^k y_{T,j,m}^k e^{in\left( \phi_{i,m}^k-\phi_{j,m}^k\right)} \right).
\end{split}
\end{equation}

\noindent Therefore,  $\left \langle \left \langle 2 \right \rangle \right \rangle$ of the azimuthal distribution of MTR is as follows:

\begin{equation}
\label{22-yt} \langle \left \langle 2 \rangle \right \rangle_{y_T} =\frac{\sum_{k=1}^{N_{\rm event}}\langle 2\rangle^k_{\rm d}+\langle 2\rangle^k_{\rm s} }{\sum_{k=1}^{N_{\rm event}}W^k},
\end{equation}

\noindent where $W^k$ is the event weight, i.e.,

\begin{equation}
\begin{split}\label{weight}
W^k = &\sum_{m,n=1,m\neq n}^{N_{\rm bin}} \frac{1}{N_m^k N_n^k} \sum_{i=1}^{N_m^k} \sum_{j=1}^{N_n^k} y_{T,i,m}^k y_{T,j,n}^k \\  \
&+\sum_{m=1}^{N_{\rm bin}}  \frac{1}{N_m^k\left ( N_m^k-1 \right )} \sum_{i,j=1,i\neq j}^{N_m^k}y_{T,i,m}^ky_{T,j,m}^k.
\end{split}
\end{equation}

\noindent The corresponding anisotropic radial flow is as follows:

\begin{equation}
\label{v22-yT}v_{2} (y_T) =\sqrt{\left \langle \left \langle 2\right \rangle \right \rangle_{y_T}}.
\end{equation}

Up to now, we derive the anisotropic coefficients of azimuthal multiplicity, TTR and MTR distributions by both the EPM and the TPCM. In order to check how these two methods work in practice, we apply them to the event sample generated by Monte Carlo simulation in the following section.

\section{Application}

\begin{figure*}[htb]
\includegraphics[width=6.5in]{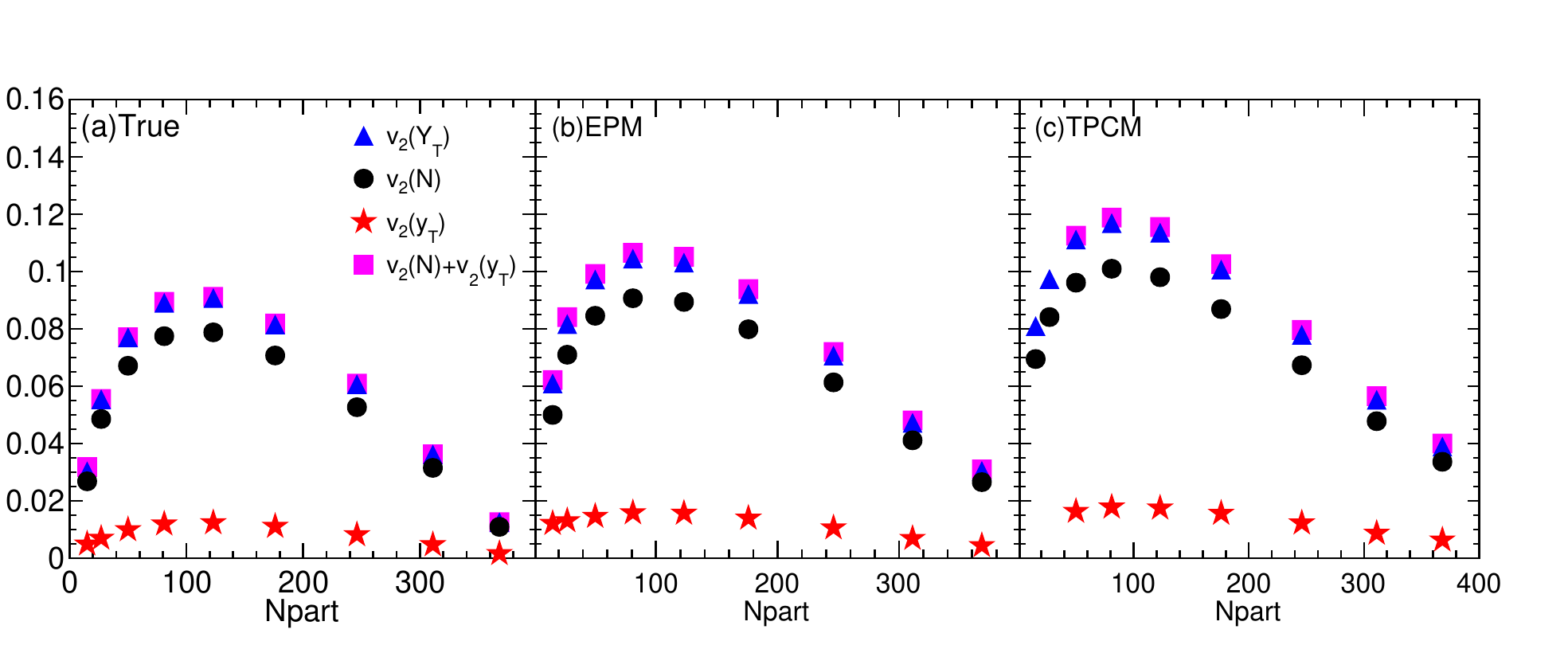}
\caption{\label{Fig. 1} Centrality dependence of elliptic flow (solid black dots), anisotropic radial flow (solid red stars), anisotropic TTR flow (solid blue triangles), and the summation of elliptic flow and anisotropic radial flow (solid purple squares). (a) is the true results. (b) and (c) are the results estimated by the EPM and the TPCM, respectively. The error of each point is smaller than the symbol size.}
\end{figure*}

The AMPT model with string melting~\cite{lab38,lab39} can reproduce the observed elliptic flow at the RHIC~\cite{lab32,lab33,lab34,lab35}. The event sample generated by this model has the basic characteristics of flows, in general, and can be used to test the effectivity of the two aforementioned methods. 

With this model, we generate $4\times10^6$ events for Au + Au collisions at 200 GeV. The analysis is presented in the kinetic ranges $|\eta |\leq 2.5$, and $p_{T} \in \left [ 0.15, 2 \right ]\rm GeV/\mathit c$ which is the same as the RHIC/STAR Collaboration~\cite{lab33,lab35}. Nine centrality bins are defined by the $N_{\rm part}$ corresponding to the nine multiplicity-ranges, consistent with those at the RHIC/STAR Collaboration~\cite{lab35}.

For the EPM, two subevents [(a) and (b)] for resolution corrections are constructed in two longitudinal symmetry windows $|\eta| \in \left [ 3.3, 4.5 \right ] $, similar to those at the RHIC/STAR Collaboration~\cite{lab23}. We also use a random method to make the number of charged particles in each subevent even~\cite{lab29,lab33}. 

Since the azimuthal angle of the true reaction plane $\Psi_{r}$ is known and set to zero in this model, $v_{2}(Y_T), v_{2}(N)$, and $v_{2}(y_{T})$ can be directly obtained by the corresponding azimuthal distributions. These directly obtained $v_{2}(Y_T), v_{2}(N)$, and $v_{2}(y_{T})$ can be regarded as their true values. Their centrality dependence are represented in Fig.~1(a) by solid blue triangles, black dots, and red stars, respectively. 

\begin{figure*}[htb]
\includegraphics[width=7.2in]{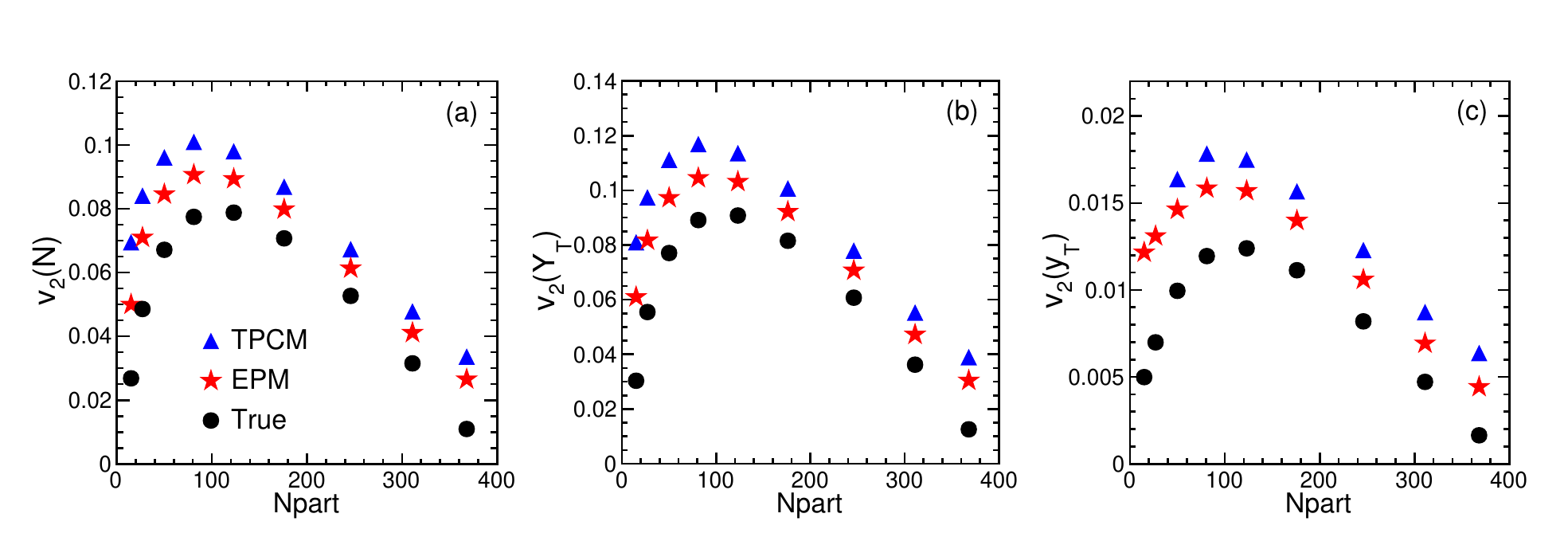}
\caption{\label{Fig. 2} Centrality dependence of (a) $v_{2}(N)$, (b) $v_{2}(Y_T)$, and (c) $v_{2}(y_{T})$ where solid black dots are true values, solid red stars, and solid blue triangles are estimated by the EPM and the TPCM, respectively. The error of each point is smaller than the symbol size.}
\end{figure*}

In Fig.~1(a), three kinds of flows have similar centrality dependence, consistent with those obtained from UrQMD~\cite{lab27,lab28,lab40}. In each of the centrality interval, $v_{2}( Y_{T})$ is the largest, $v_{2}(N)$ is medial, and  $v_{2} (y_{T})$ is the smallest. It is no surprise that $v_{2}(N)$ and $v_{2} (y_{T})$ are both smaller than $v_{2}( Y_{T})$. From the definitions of Eqs.~(\ref{YT}), (\ref{yT}), and (\ref{N}), TTR contains the contributions from both the number of particles (multiplicity) and the transverse movement. MTR is raw kinetics, and the influence of multiplicity is excluded. Therefore, $v_{2} (y_{T})$ and $v_{2}(N)$ are both smaller than $v_{2}( Y_{T})$. 

It is interesting that $v_{2} (y_{T})+v_{2}(N)$ (purple squares) overlaps with $v_{2}( Y_{T})$ (blue triangles) at each of the nine centralities. This relationship shows that the anisotropic radial flow and elliptic flow are additive, although MTR and multiplicity are not simply related to TTR.

In order to compare the anisotropic coefficients estimated by the two methods with their true values of $v_{2}(N), v_{2}(Y_T)$, and $v_{2}(y_T)$ are presented in Figs.~2(a)-2(c), respectively, where solid black dots are the true values, solid red stars, and blue triangles are estimated by the EPM and the TPCM.

For $v_{2}(N)$ as shown in Fig.~2(a), in each centrality interval, the black dot is the lowest, the red star is in the middle, and the blue triangle is the highest. This shows that the estimation given by TPCM is higher that given by the EPM, and the estimations given these two methods are both higher than the true value. The excess parts are the contributions of nonflow and flow fluctuations~\cite{lab41,lab42}. Nevertheless, the difference is roughly an overall shift. The general trends of three data sets are consistent with each other and with the expectation that the larger elliptic flow appears in the midcentral collisions and small elliptic flow in peripheral and central collisions. Therefore, both methods can be considered as a valid estimation of true elliptic flow.

Similarly, for $v_{2}(Y_{T})$ as shown in Fig.~2(b), in each centrality interval, three kinds of points have the same position ordering as shown in Fig.~2(a), i.e., the lowest black dot, the middle red star, and the highest blue triangle. By the same reasoning, these results demonstrate that the both methods are effective in estimating $v_{2}(Y_{T})$. 

For $v_{2}(y_{T})$ as shown in Fig.~2(c), three kinds of points also display the same ordering as shown in Figs.~2(a) and 2(b), i.e., the lowest black dots, the middle red stars, and the highest blue triangles for the seven middle and central collisions. These results are not really surprising. As even in some toy Monte Carlo studies where there is only flow fluctuations and no systematic biases due to nonflow, the elliptic flows given by the EPM and the TPCM are still different~\cite{lab42}.
 
The centrality dependence of elliptic flow, anisotropic TTR flow, and radial flow estimated by the EPM and the TPCM are presented in Figs.~1(b) and 1(c), respectively, in comparison to their true values in Fig.~1(a). The summations of elliptic flow $v_{2}(N)$ and anisotropic radial flow $v_{2} (y_{T})$ (purple squares) estimated by these two methods also overlap with corresponding anisotropic TTR flow $v_{2} (Y_{T})$ (blue triangles), consistent with the case in Fig.~1(a). Therefore, the elliptic flow and anisotropic radial flow estimated by these two methods are also additive, just like their true values. This demonstrates that the two methods are equally effective.
 
However, the TPCM is ineffective in estimating $v_2(y_T)$ for the two peripheral-collision bins. We will show in the following section why and how the TPCM fails.

\section{The limitation of the TPCM}

$v_2(y_T)$ given by the TPCM in Eq.~(\ref{v22-yT}) is the square root of $\langle \langle 2\rangle\rangle_{y_T}$. For the two peripheral-collision bins, $\langle \langle 2\rangle\rangle_{y_T}$ becomes negative. This is why the TPCM fails.

We can understand the cause of this problem from the original definition of MTR. MTR is defined to be averaged over $N_m$ (the number of particles in the \emph{m}th bin). If all particles are uniformly distributed in the whole azimuthal region $N_m\sim N/N_{\rm bin}$ ($N$ is multiplicity, and $N_{\rm bin}$ is the total number of bins). Now, assuming $\langle \langle 2\rangle\rangle_{y_T}$ is well defined, we should have selected a proper $N_{\rm bin}$ for which $N_m$ is large enough. 




To examine the proper range of $N_{\rm bin}, \langle \langle 2\rangle\rangle_{y_T}$ versus $N_{\rm bin}$ at the nine centralities are presented in Fig.~3. For midcentral collisions and central collisions, $\langle \langle 2\rangle\rangle_{y_T}$ decreases with $N_{\rm bin}$ when $N_{\rm bin}\leq 20$, and then becomes independent of $ N_{\rm bin}$ when $N_{\rm bin} \geq 20 $. There is a wide range of $N_{\rm bin}$, where $\langle \langle 2\rangle\rangle_{y_T}$ stays flat. This range guarantees that $\langle \langle 2\rangle\rangle_{y_T}$ is independent of $N_{\rm bin}$. Therefore, we choose $N_{\rm bin}=25$ for all the aforementioned analysis. 

\begin{figure}[htb]
\includegraphics[width=3.3in]{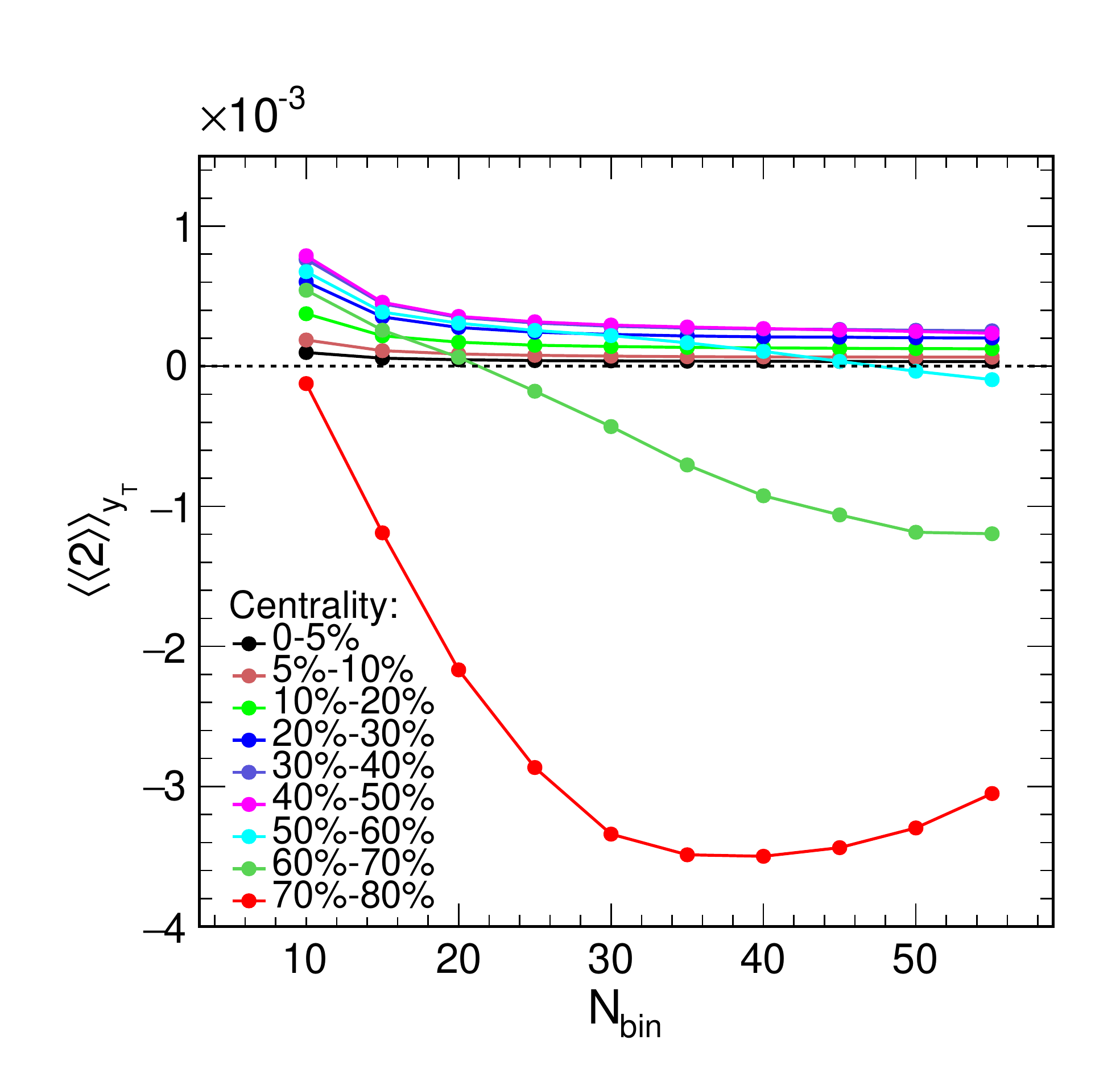}
\caption{\label{Fig. 3}  $\langle \langle 2\rangle\rangle_{y_T}$ versus $N_{\rm bin}$ at the nine centralities where the error of each point is smaller than the symbol size.}
\end{figure}

However, for the two peripheral-collision bins, $\langle \langle 2\rangle\rangle_{y_T}$ rapidly decreases to negative as $N_{\rm bin}$ increases. Obviously, this is not caused by the change in $N_{\rm bin}$ but instead by the decrease of multiplicity $N$ from central to peripheral collisions. When $N_{\rm bin}$ is fixed, $N_m$ is solely proportional to $N$. Therefore, $N_m$ also decreases from central to peripheral collisions.



To find the lower threshold of $\la N_m\ra $ in general, we randomly drop some particles in each event. Using this method, we can show how $\langle \langle 2\rangle\rangle_{y_T}$ changes with $\la N_m\ra$, and where in each centrality interval the TPCM becomes invalid. 

$\langle \langle 2\rangle\rangle_{y_T}$ versus $\la N_m\ra $ for four cases are presented in Figs.~4(a)-4(d). For the two upper subfigures, we choose the pseudorapidity range $|\eta|\leq2.5$ and $N_{\rm bin}=25$ for Fig.~4(a) and $N_{\rm bin}=35$ for Fig.~4(b). In Fig.~4(a),  with a decrease in $\la N_m \ra, \langle \langle 2\rangle\rangle_{y_T}$ stays flat and positive. This implies that $\langle \langle 2\rangle\rangle_{y_T}$ can still be estimated correctly, even though some particles in each of the events are dropped. When $\la N_m\ra$ decreases even lower, $\langle \langle 2\rangle\rangle_{y_T}$ rapidly drops to negative. This implies that the correlations between the two particles decrease significantly, and the method becomes invalid. Meanwhile, Fig.~4(b) also shows the same $\la N_m\ra$ which makes $\langle \langle 2\rangle\rangle_{y_T}$ rapidly drop  to negative. Therefore, independent of $N_{\rm bin}$, as long as the mean number of particles in a bin is less than 10, the TPCM will fail.

\begin{figure}[htb]
\includegraphics[width=3.4in]{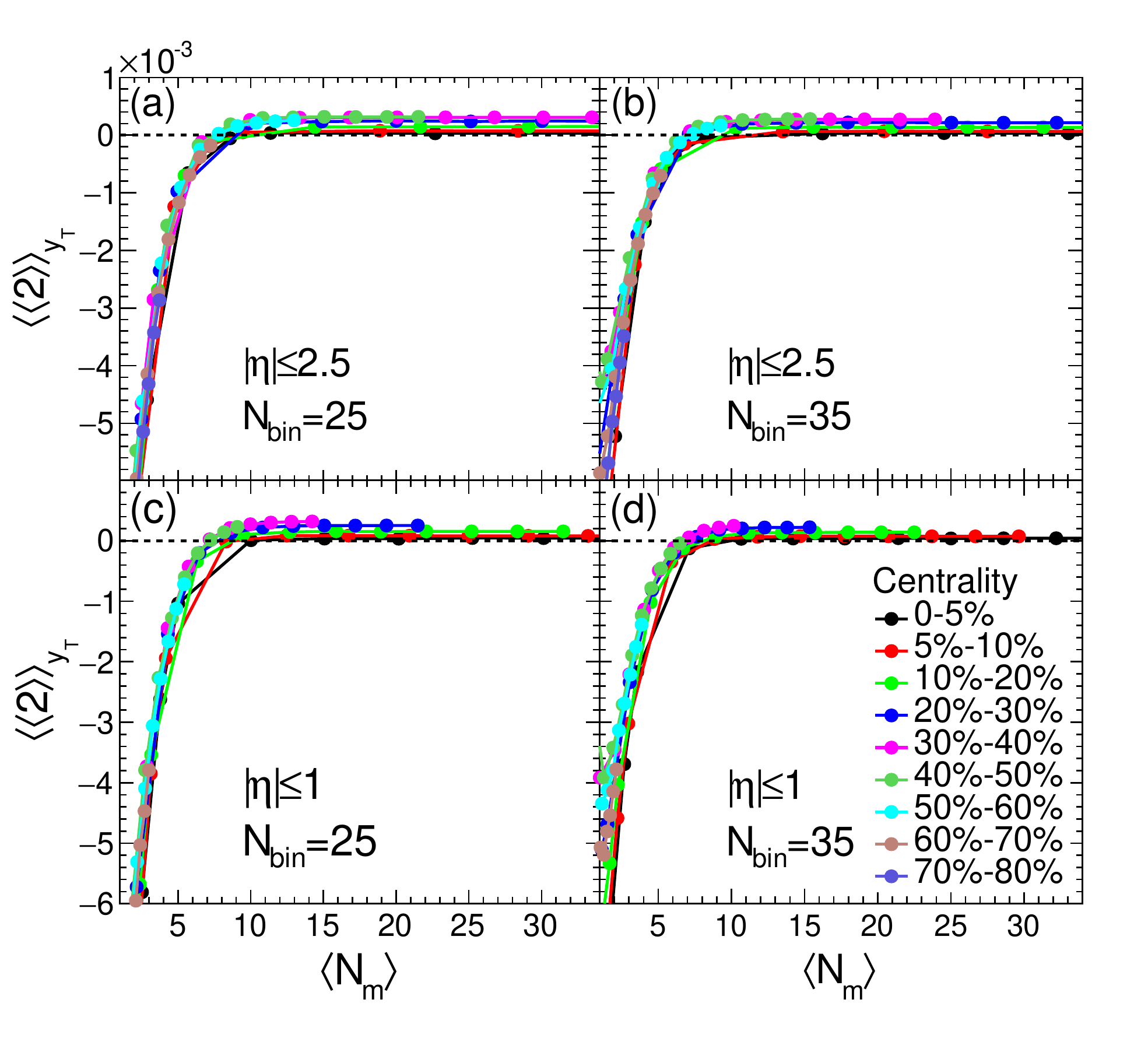}
\caption{\label{Fig. 4}  $\langle \langle 2\rangle\rangle_{y_T}$ versus $\la N_m\ra $ at the nine centralities where the error of each point is smaller than the symbol size.}
\end{figure}

For the two lower subfigures, we choose the pseudorapidity range $|\eta|\leq1$ and $N_{\rm bin}=25$ for Fig.~4(c) and $N_{\rm bin}=35$ for Fig.~4(d). In contrast with the two upper row subfigures, narrow pseudorapidity range makes the total multiplicity smaller. The general trends in Figs.~4(c) and 4(d) are consistent with each other and with the two upper subfigures. This means that $\langle \langle 2\rangle\rangle_{y_T}$ is independent of the total mean multiplicity $\la N\ra$ and $N_{\rm bin}$ but is dependent on $\la N_m\ra$.  Both Figs.~4(c) and 4(d) show negative $\langle \langle 2\rangle\rangle_{y_T}$ when $\la N_m\ra <10$. Therefore, the lower threshold of $\la N_m\ra $  is ten for the TPCM.

In addition, one may ask why there is no such lower threshold in the EPM. This is because $v_2(y_T)$ given by Eq.~(\ref{v22-yT}) is of higher order than that given by Eq.~(\ref{v2-yT}). The right side of Eq.~(\ref{v22-yT}) is the square root of two-particle correlations. Therefore, higher precision is required for the TPCM.

Moreover, as shown in Fig.~1, anisotropic radial flow is the smallest one among three. Its measurement is more difficult than anisotropic TTR flow, or elliptic flow. Therefore, a larger number of two-particle pairs is required. 

\section{Summary and conclusions}

Anisotropic radial flow can be directly extracted from the azimuthal distribution of the MTR. First, we apply the EPM and the TPCM to estimate the anisotropic coefficients of the azimuthal distributions of TTR and MTR in relativistic heavy-ion collisions. 

Then, using the event sample of Au + Au collisions at 200 GeV generated by the AMPT model with string melting, we show that the EPM and the TPCM are both effective in estimating the anisotropic TTR flow. The EPM is also effective in estimating anisotropic radial flow. 

For central collisions and midcentral collisions, the TPCM is effective in estimating anisotropic radial flow. However, for the two peripheral-collision bins, this method fails when the mean number of particles in a bin becomes less than ten. 

Therefore, anisotropic radial flow can be estimated in a model-independent way by the EPM and the TPCM in relativistic heavy-ion collisions. For the TPCM to be reliable, the mean number of particles in an azimuthal bin must be above a certain threshold.

We find that the summation of the true elliptic flow and true anisotropic radial flow is consistent with the true anisotropic TTR flow. This relationship also holds for the corresponding coefficients estimated by the EPM and the TPCM. Therefore, elliptic flow and anisotropic radial flow are additive.

\section{Acknowledgement}

We are very grateful to Dr. G. Wang for his valuable comments and suggestions. We thank Dr. A. Tang, F. Wang, Y. Zhou, and Z. Lin for helpful discussions. This work was supported, in part, by the Ministry of Science and Technology (MoST) under Grant No. 2016YFE0104800 and the Fundamental Research Funds for the Central Universities under Grant No. CCNU19ZN019.

\ed
\begin{thebibliography}{99}

\bibitem{lab1} J. Y. Ollitrault, Nucl. Phys. A 638, 195c (1998).

\bibitem{lab2} P. Braun-Munzinger, J. Stachel, Nature  (London) 448, 302 (2007).

\bibitem{lab3} S. A. Voloshin, Phys. Rev. C 55, R1630(R) (1997).

\bibitem{lab4} S. A. Voloshin, A. M. Poskanzer, and R. Snellings, in \emph{Landolt-Boernstein, Relativistic Heavy Ion Physics}, Vol. 1/23 (Springer-Verlag, 2010), pp. 5–54.

\bibitem{lab5} T. Hirano, arXiv:nucl-th/9904082.

\bibitem{lab6} P. Danielewicz, Phys. Rev. C 51, 716 (1995).

\bibitem{lab7} C. M. Hung and E. Shuryak, Phys. Rev. C 57, 1891 (1998).

\bibitem{lab8} P. Huovinen and P. V. Ruuskanen, Annu. Rev. Nucl. Part. Sci. 56, 163 (2006).

\bibitem{lab9} M. Shao, L. Yi, Z. B. Tang, H. F. Chen \emph{et al.}, J. Phys. G: Nucl. Part. Phys. 37, 085104 (2010).

\bibitem{lab10} Z. B. Tang, L. Yi, L. J. Ruan \emph{et al.}, Chin. Phys. Lett. 30, 031201 (2013).

\bibitem{lab11} J. Chen, J. Deng, Z. Tang, Z. Xu, and L. Yi, arXiv: 2012.02986.

\bibitem{lab12} L. Li, N. Li, and Y. F. Wu, J. Phys. G: Nucl. Part. Phys. 40, 075104 (2013).

\bibitem{lab13} P. Huovinen, P. F. Koblb, U. W. Heinz, P. V. Ruuskanen, and S. A. Voloshin, Phys. Lett. B 503, 58 (2001).

\bibitem{lab14} N. Borghini and J. Y. Ollitrault, Phys. Lett. B 642, 227 (2006).

\bibitem{lab15} L. Li, N. Li, and Y. F. Wu, Chin. Phys. C 36, 423 (2012).

\bibitem{lab16} L. D. Landau and E. M. Lifshitz, \emph{Fluid Mechanics}, 2nd. ed., Vol. 6, Course of Theoretical Physics (Pergamon, Oxford, 1987), p.44.

\bibitem{lab17} C. Shen and U. Heinz, Phys. Rev. C 85, 054902 (2012).

\bibitem{lab18} Iu. A. Karpenko, P. Huovinen, H. Petersen, and M. Bleicher, Phys. Rev. C 91, 064901 (2015).

\bibitem{lab19} P. Liu and R. A. Lacey, Phys. Rev. C 98, 021902 (2018).

\bibitem{lab20} E. Schnedermann, J. Sollfrank, and U. W. Heinz, Phys. Rev. C 48, 2462 (1993).

\bibitem{lab21} Y. Oh, Z. W. Lin, and C. M. Ko, Phys. Rev. C 80, 064902 (2009).

\bibitem{lab22} J. Barrette \emph{et al.} (E877 Collaboration), Phys. Rev. C 62, 024901 (2000).

\bibitem{lab23} L. Adamczyk \emph{et al.} (STAR Collaboration), Phys. Rev. C 93, 014907 (2016).

\bibitem{lab24} L. Adamczyk \emph{et al.} (STAR Collaboration), Phys. Rev. C 96, 044904 (2017).

\bibitem{lab25} B. Abelev \emph{et al.} (ALICE Collaboration), Phys. Rev. C 88, 044910 (2013).

\bibitem{lab26} P. Yang, L. Li, and Y. F. Wu, Beijing: Sciencepaper Online, 201404-454 (2014) [arXiv:1405.0686].

\bibitem{lab27} S. Sarkar, P. Mali, and A. Mukhopadhyay, Phys. Rev. C 95, 014908 (2017).

\bibitem{lab28} S. Sarkar, P. Mali, S. Ghosh, and A. Mukhopadhyay, Adv. High Energy Phys. 2018, 7453752 (2018).

\bibitem{lab29} A. M. Poskanzer and S. A. Voloshin, Phys. Rev. C 58, 1671 (1998).

\bibitem{lab30} A. Bilandzic, R. Snellings, and S. A. Voloshin, Phys. Rev. C 83, 044913 (2011).

\bibitem{lab31} N. Borghini, P. M. Dinh, and J. Y. Ollitrault, Phys. Rev. C 64, 054901 (2001).

\bibitem{lab32} C. Adler \emph{et al.} (STAR Collaboration), Phys. Rev. C 66, 034904 (2002).

\bibitem{lab33} B. I. Abelev \emph{et al.} (STAR Collaboration), Phys. Rev. C 77, 054901 (2008).

\bibitem{lab34} L. Adamczyk \emph{et al.} (STAR Collaboration), Phys. Rev. C 86, 054908 (2012).

\bibitem{lab35} J. Adams \emph{et al.} (STAR Collaboration), Phys. Rev. C 72, 014904 (2005).

\bibitem{lab36} S. Wang, Y. Z. Jiang, Y. M. Liu, D. Keane, D. Beavis, S. Y. Chu, S. Y. Fung, M. Vient, C. Hartnack, and H. Stöcker, Phys. Rev. C 44, 1091 (1991).

\bibitem{lab37} J. Y. Ollitrault, Phys. Rev. D 48, 1132 (1993).

\bibitem{lab38} Z. W. Lin and C. M. Ko, Phys. Rev. C 65, 034904 (2002).

\bibitem{lab39} Z. W. Lin, C. M. Ko, B. A. Li, B. Zhang and S. Pal, Phys. Rev. C 72, 064901 (2005).

\bibitem{lab40} S. Sarkar and A. Mukhopadhyay, Proc. DAE-BRNS Symp. Nucl. Phys. 60, 752 (2015).

\bibitem{lab41} J. Y. Ollitrault, A. M. Poskanzer and S. A. Voloshin, Phys. Rev. C 80, 014904 (2009).

\bibitem{lab42} M. Luzum and J. Y. Ollitrault, Phys. Rev. C 87, 044907 (2013).


\end{thebibliography}
